\documentclass[sigconf]{acmart}

\copyrightyear{2022}
\acmYear{2022}
\setcopyright{rightsretained}
\acmConference[MAD '22]{Proceedings of the 1st International Workshop on Multimedia AI against Disinformation}{June 27--30, 2022}{Newark, NJ, USA}
\acmBooktitle{Proceedings of the 1st International Workshop on Multimedia AI against Disinformation (MAD '22), June 27--30, 2022, Newark, NJ, USA}
\acmDOI{10.1145/3512732.3533586}
\acmISBN{978-1-4503-9242-6/22/06}

\usepackage{etoolbox}
\makeatletter
\patchcmd{\maketitle}{\@copyrightpermission}{
   \begin{minipage}{0.3\columnwidth}
     \href{http://creativecommons.org/licenses/by/4.0/}{\includegraphics[width=0.90\textwidth]{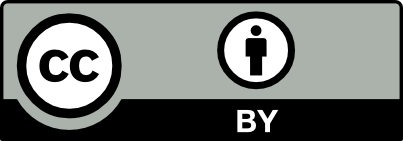}}
   \end{minipage}\hfill
   \begin{minipage}{0.7\columnwidth}
     \href{http://creativecommons.org/licenses/by/4.0/}{This work is licensed under a Creative Commons Attribution International 4.0 License.}
   \end{minipage}
 
   \vspace{5pt}
}{}{}

\makeatother

\DeclareMathOperator{\denoise}{denoise}
\DeclareMathOperator{\classify}{classify}
\newcommand{\Nfft}{N_{\text{stft}}}
\newcommand{\posterior}[2]{\mathcal{N}\left( #1 \mid \mu_{#2}, \Sigma_{#2} \right)}
\DeclareMathOperator{\sgn}{sgn}
\newcommand{\lnorm}[1]{ \left\lVert #1 \right\rVert }
\newcommand{\abs}[1]{ \left\lvert #1 \right\rvert }

\usepackage{cleveref}  
\usepackage{glossaries}
\usepackage{booktabs, multirow, multicol}
\usepackage{graphicx}
\usepackage{subcaption}
\glsdisablehyper

\newcommand{\specgramminipagewidth}{0.25\linewidth}
\newcommand{\specgramwidth}{0.9\linewidth}

\newacronym{ai}{AI}{Artificial Intelligence}
\newacronym{dsp}{DSP}{Digital Signal Processing}
\newacronym{gmm}{GMM}{Gaussian Mixture Model}
\newacronym{lstm}{LSTM}{Long Short-Term Memory}
\newacronym{mfcc}{MFCC}{Mel-frequency cepstral coefficient}
\newacronym{snr}{SNR}{Signal-to-Noise Ratio}
\newacronym{stft}{STFT}{Short Time Fourier Transform}
\newacronym{svm}{SVM}{Support Vector Machine}
\newacronym{rbf}{RBF}{Radial Basis Function}

\newacronym{psnr}{PSNR}{Peak Signal-to-Noise Ratio}
\newacronym{ssim}{SSIM}{Structural Similarity Index Measure}
\newacronym{mca}{MCA}{Microphone Classification Accuracy}

\begin{document}
\fancyhead{}

%
%

\title{Spectral Denoising for Microphone Classification}

\author{Luca Cuccovillo}
\orcid{0000-0001-5559-6508}
\affiliation{%
 \institution{Fraunhofer IDMT}
 \streetaddress{Ehrenbergstr. 31}
 \city{Ilmenau}
 \country{Germany}}
\email{luca.cuccovillo@idmt.fraunhofer.de}

\author{Antonio Giganti}
\orcid{0000-0003-4052-5138}
\affiliation{%
  \institution{Politecnico di Milano}
  \streetaddress{P.zza Leonardo da Vinci 32}
  \city{Milano}
  \country{Italy}}
\email{antonio.giganti@polimi.it}

\author{Paolo Bestagini}
\orcid{0000-0003-0406-0222}
\affiliation{%
  \institution{Politecnico di Milano}
  \streetaddress{P.zza Leonardo da Vinci 32}
  \city{Milano}
  \country{Italy}}
\email{paolo.bestagini@polimi.it}

\author{Patrick Aichroth}
\orcid{0000-0003-4777-6335}
\affiliation{%
 \institution{Fraunhofer IDMT}
 \streetaddress{Ehrenbergstr. 31}
 \city{Ilmenau}
 \country{Germany}}
\email{patrick.aichroth@idmt.fraunhofer.de}

\author{Stefano Tubaro}
\orcid{0000-0002-1990-9869}
\affiliation{%
  \institution{Politecnico di Milano}
  \streetaddress{P.zza Leonardo da Vinci 32}
  \city{Milano}
  \country{Italy}}
\email{stefano.tubaro@polimi.it}



\begin{abstract}
  In this paper, we propose the use of denoising for microphone classification, to enable its usage for several key application domains that involve noisy conditions. We describe the proposed analysis pipeline and the baseline algorithm for microphone classification, and discuss  various denoising approaches which can be applied to it within the time or spectral domain; finally, we determine the best-performing denoising procedure, and evaluate the performance of the overall, integrated approach with several SNR levels of additive input noise. As a result, the proposed method achieves an average accuracy increase of about 25\% on denoised content over the reference baseline.
\end{abstract}

\ccsdesc[500]{Security and privacy}
\ccsdesc[300]{Information systems~Multimedia information systems}
\ccsdesc[300]{Information systems~Social networks}
\ccsdesc[500]{Computing methodologies~Artificial intelligence}
\ccsdesc[300]{Computing methodologies~Machine learning}

\keywords{%
Microphone classification,
Audio forensics,
Spectral denoising,
AI-based denoising,
Digital signal processing,
Machine learning%
}


\maketitle

%
%

\section{Introduction}

The recent surge of disinformation, which often includes malevolent manipulation, decontextualization or fabrication of audio-visual material via social media, has drawn increasing attention from the research community, which reacted by developing innovative algorithms for detecting frauds and analysing controversial multimedia content~\cite{reverse_video_search,twitter_analysis}. In particular, the discipline of  multimedia forensics has found a new momentum: The need for tools to analyze acquisition and processing traces within content, and (i) to compare them with alleged information about the content~\cite{openset_microphone_classification,openset_camera_classification}  and (ii) to use it to detect and localize manipulations~\cite{mic_class:cuccovillo,video_tampering_detection}. Such tools are crucial to fight media disinformation, both in investigative journalism, and in courtroom cases.

In this paper, we address an important challenge related to microphone classification, a classic task within the forensics domain that aims at identifying which device has been used to record a given audio item~\cite{mic_class:cuccovillo,openset_microphone_classification,mic_class:kraetzer,mic_class:lin,mic_class:luo}: Until now, microphone classification algorithms have focused on the analysis of fairly high-quality content, which is common e.g. in courtroom cases. They are therefore very sensitive to background or additive noise. This sensitivity to noise, however, is greatly reducing the applicability of microphone classification to the kind of audio-visual disinformation that is increasingly shared via social media, which often includes noisy audio material.

Considering this, the motivation for this paper was to investigate whether and which denoising techniques based either on \gls{dsp} or \gls{ai} could be applied as a pre-processing step for microphone classification, to improve its robustness against noise. Moreover, we propose an integrated algorithm for microphone classification using such denoising: It is is based on our own pre-existing baseline for closed-set microphone classification~\cite{mic_class:cuccovillo}, later extended also to address an open-set setup~\cite{openset_microphone_classification}, and uses the best-performing denoising algorithm we were able to identify within this work: the \gls{ai}-based DnCNN architecture for image denoising proposed by \citeauthor{denoiser:dncnn}~\cite{denoiser:dncnn}, which we applied to the spectral domain. This new application of AI to spectral denoising for microphone classification achieved very promising results, with an average accuracy increase of about 25\% in comparison to the baseline. 

The further chapters of the paper are organized as follows: In \Cref{sec:proposed-approach}, we present our proposed integrated approach for microphone classification, based on the baseline algorithm that is described in \Cref{sec:micclass-baseline}. In \Cref{sec:denoising-baselines}, we present various denoising techniques, which are then compared in \Cref{sec:denoiser-selection} to select the most suitable one for microphone classification. Finally, in \Cref{sec:evaluation}, we evaluate the proposed integrated approach with varying levels of additive noise. \Cref{sec:outlook} closes with a a discussion about future research directions and ideas for further improvements.
\section{Proposed Approach}\label{sec:proposed-approach}
\begin{figure}[ht]
    \centering
    \includegraphics[width=\columnwidth,trim={2cm 5.75cm 2cm 8cm},clip]{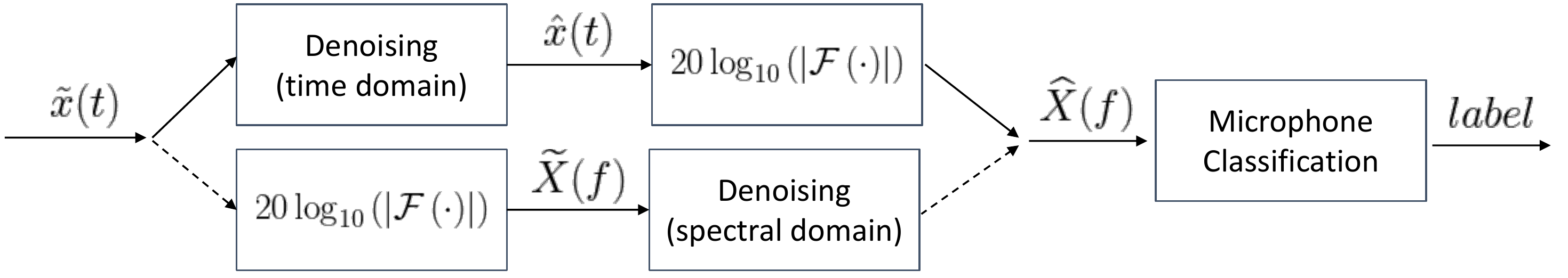}
    \caption{Overview of the integrated approach}
    \label{fig:overview}
\end{figure}

Our integrated approach for microphone classification consists of three main components:
\begin{enumerate}
    \item A \emph{denoising} component, which analyses a noisy input signal $\tilde{x}$ and returns an estimate $\hat{x}$ of the input signal $x$ before being affected by any additive noise
    \item A \emph{log-power extraction} component, which takes an input signal in the time domain $x(t)$ and returns its logarithmic power spectrum $X(f)$
    \item A \emph{microphone classification} component, which uses an input power spectrum $X(f)$ to return the $label$ of the microphone device used for recording the initial audio signal
\end{enumerate}
As depicted in \Cref{fig:overview}, the order of the described steps depends on the target domain of the denoising component. 

If the denoising component is designed to work on a signal $x(t)$ in the time domain, then denoising is applied first, to be followed by log-power extraction. The denoised log-power is then used by microphone classification. This process is described more formally with \cref{eq:approach:time:1,eq:approach:time:2,eq:approach:time:3}
\begin{subequations}
\begin{align}
   & \hat{x}(t) = \denoise\left( \tilde{x}(t)\right) \approx x(t) \label{eq:approach:time:1}, \\
   & \widehat{X}(f) = 20\log_{10}\left(\left\lvert \mathcal{F} (x(t)) \right\rvert\right)
    \label{eq:approach:time:2}, \\
    & label =\classify\left(\widehat{X}(f)\right) \label{eq:approach:time:3},
\end{align}
\end{subequations}
where $\mathcal{F}(\cdot)$ denotes the Fourier transform, and $x(t)$ the original signal in the time domain before the noise addition. 

If the denoising component is designed to work on a signal $X(f)$ in the frequency domain, then log-power extraction is performed first, followed by denoising that is applied as desired on the frequency domain. Again, the denoised log-power is then used by microphone classification. This process is described more formally with \cref{eq:approach:frequency:1,eq:approach:frequency:2,eq:approach:frequency:3}
\begin{subequations}
\begin{align}
   & \widetilde{X}(f) = 20\log_{10}\left(\left\lvert \mathcal{F} (\tilde{x}(t)) \right\rvert\right)
    \label{eq:approach:frequency:1}, \\
   & \widehat{X}(f) = \denoise\left( \widetilde{X}(f) \right) \approx X(f) \label{eq:approach:frequency:2}, \\
    & label =\classify\left(\widehat{X}(f)\right) \label{eq:approach:frequency:3},
\end{align}
\end{subequations}
where $X(f)$ denotes the original signal in the frequency domain before the noise addition. 

%
%

The denoising components in the frequency domain that can be applied to the integrated approach are less constrained than ``classic'' ones: Since we are not going to convert the denoised signal $\widehat{X}(f)$ back into the time domain, candidate algorithms can ignore the phase of the signal completely, focusing only on its log-power. Moreover, we do not need to work on the whole signal length at once, but can rather focus on a set of $L$ analysis frames, each of which being denoted by $x_l(t)$. 
An important consequence of these relaxed constraints is that if we consider the whole \gls{stft} of the signal at once, i.e., a set of $L$ frames in which the log-power of the $l$-th frame can be denoted by $X_l(f)$, the denoising operation can be performed also by means of \emph{image denoising algorithms}, which have been thoroughly investigated~\cite{denoiser:total_variation, denoiser:nonlocal_means, denoiser:bilateral_filtering, denoiser:wavelet_bayesian_shrink, denoiser:dncnn}.


\section{Microphone Classification Baseline}\label{sec:micclass-baseline}
The baseline approach for microphone classification consists of our own previous method that is based on blind channel estimation~\cite{mic_class:cuccovillo,openset_microphone_classification}.

\subsection{Channel Estimation}

The algorithm starts from the assumption that each frame $x_l(t)$ of an input recording $x(t)$ can be modeled by a convolution between a \emph{fixed} transmission channel $h(t)$, and the original input speech $s_l(t)$, i.e.:
\begin{equation}
x_l(t)=h(t)\ast s_l(t), 
\end{equation}
where we assume the transmission channel equal to the \emph{frequency response} of the recording device. 

An equivalent formulation in the log-power domain is:
\begin{equation}
X_l(f)=H(f) + S_l(f),
\end{equation}
which provides us with a straight-forward solution to estimate the log-power $H(f)$ of the transmission channel: If we can estimate the ideal input speech $S_l(f)$, i.e., if we can compute a term $\widehat{S}_l(f)$ which is accurate enough, the channel can be estimated blindly by applying: 
\begin{equation}
\widehat{H}(f)=\frac{1}{L}\sum_{l=1}^L \left(X_l(f)-\widehat{S}_l(f)\right),\label{eq:channel-estimation:pre-final}
\end{equation}
with $L$ denoting the amount of frames and thus \cref{eq:channel-estimation:pre-final} denoting the average difference between the input recording frames and the estimated ideal speech. 

To further improve the stability of \cref{eq:channel-estimation:pre-final}, we can normalize both terms $X_l(f)$ and $\widehat{S}_l(f)$ to have zero mean, by defining their normalized equivalents
\begin{subequations}
\begin{align}
   Z_{X_l}(f) & =  X_l(f) - \frac{1}{\Nfft}\sum_{f=1}^{\Nfft} X_l(f),\\
   Z_{\widehat{S}_l}(f) & = \widehat{S}_l(f) - \frac{1}{\Nfft}\sum_{f=1}^{\Nfft} \widehat{S}_l(f),
\end{align}
\end{subequations}
and obtain the final \cref{eq:channel-estimation} for the mean-normalized blind estimation of the microphone frequency response: 
\begin{equation}
\hat{h}=\frac{1}{L}\sum_{l=1}^L \left(Z_{X_l}(f)-Z_{\widehat{S}_l}(f)\right).\quad \label{eq:channel-estimation}
\end{equation}

\subsection{Ideal Speech Estimation}
The ideal speech estimate $\widehat{S}_l(f)$ in \cref{eq:channel-estimation:pre-final} can be retrieved by leveraging spectrum classification, as firstly proposed by \citeauthor{Gaubitch:channel}~\cite{Gaubitch:channel}.

The first step for of the procedure consists of processing a large speech corpus to extract a high amount of RASTA filtered \glspl{mfcc}~\cite{RASTA_filter}, i.e., \glspl{mfcc} which are more robust to channel effects than the usual formulation, and are suited to represent phonemes. In the following, \glspl{mfcc} of the $l$-th frame of an input audio signal $x$ will be denoted by the symbol $c_{X_l}$. 

Given $L_X$ training \gls{mfcc} vectors $c_{X_l}$, used to fit a \gls{gmm}  with $M$ mixtures, a key element of the estimation procedure is the relative mixture probability $p_i\left(c_{X_l}\right)$, i.e., the probability that the feature vector $c_{X_l}$ belongs to the $i$-th mixture:
\begin{equation}
    p_i\left(c_{X_l}\right)= \frac{
      \pi_i \cdot \posterior{c_{X_l}}{i}
    }{
      \sum_{m=1}^M \pi_m \cdot \posterior{c_{X_l}}{m}
    }.\label{eq:relative_prob}
\end{equation}
In \cref{eq:relative_prob}, $\posterior{c_{X_l}}{i}$ denotes the posterior probability of the vector $c_{X_l}$ against the $i$-th mixture, having a normal distribution with mean $\mu_i$, diagonal covariance $\Sigma_i$, and prior $\pi_i$.

\begin{figure}
    \centering
    \includegraphics[width=\columnwidth]{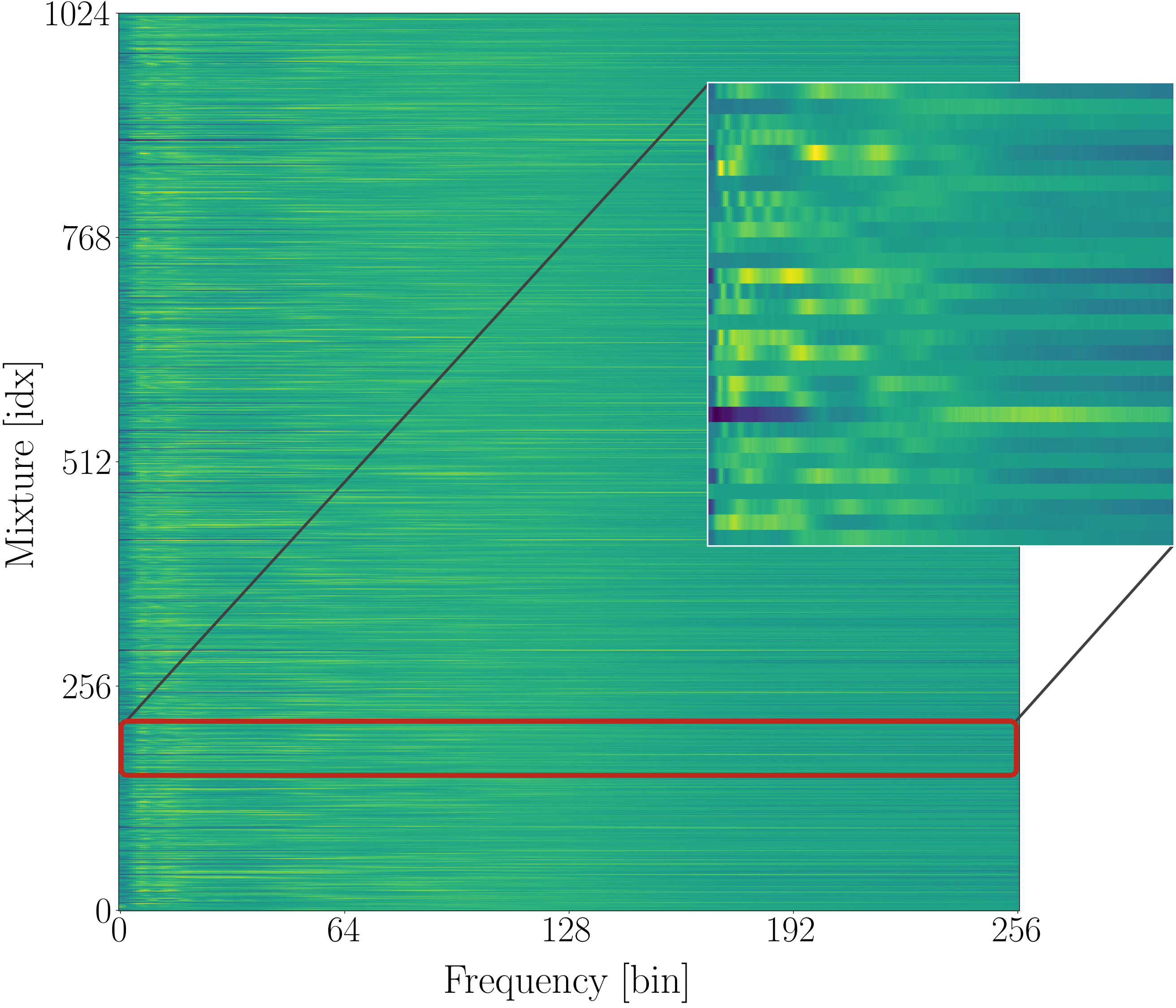}
    \caption{Average Speech Spectrum Matrix $\overline S_{X}$}
    \label{fig:average-speech-spectrum}
\end{figure}

Given these definitions, a model of the average log spectrum of the ideal speech can be obtained as follows: 
\begin{subequations}
\begin{enumerate}
    \item Build a first normalized power spectrum matrix $\overline Z_{X}$, by collecting all mean-normalized log powers $Z_{X_l}(f)$ of the \gls{gmm} training set:
    \begin{equation}
        \overline Z_{X}\in\mathbb{R}^{L_X\times\Nfft} = \left\lbrace Z_{X_l}(f) \right\rbrace,
    \end{equation}
    \item Build a relative probability matrix $\overline P_{X}$, by collecting all relative mixture probabilities $p_i\left(c_{X_l}\right)$ of the \gls{gmm} training set:
    \begin{equation}
        \overline P_{X}\in\mathbb{R}^{L_X\times M} = \left\lbrace p_i\left(c_{X_l}\right) \right\rbrace\label{eq:relative_probability_matrix}
    \end{equation}
    \item Compute the average speech spectrum matrix $\overline S_{X}$:
    \begin{equation}
        \overline S_{X}\in\mathbb{R}^{M\times \Nfft} = \overline P_{X}^t \cdot \overline Z_{X},
    \end{equation}
    with $t$ denoting the transposition.
\end{enumerate}
\end{subequations}
$\overline S_{X}$, also depicted in \Cref{fig:average-speech-spectrum}, is at the core of the ideal speech estimation procedure: Given an arbitrary input speech signal $s$ having $L_S$ frames and a relative probability matrix $\overline P_{S}$, it is straightforward to compute: 
\begin{equation}
    \overline Z_{\widehat{S}} \in \mathbb{R}^{L_S\times\Nfft} = \overline P_{S} \cdot \overline S_{X},\label{eq:ideal-speech-estimation}
\end{equation} 
i.e., a matrix the columns of which can be directly applied in  \cref{eq:channel-estimation}, to obtain the desired estimate of the microphone frequency response. 

If we analyse \cref{eq:ideal-speech-estimation} in more detail, we can observe that the relative probability matrix $\overline P_{S}$ acts as a selection matrix for the rows of the average speech spectrum $\overline S_{X}$. In other words, the ideal speech estimate $\widehat{S}_l(f)$ is obtained by means of a convex combination of the rows of $\overline S_{X}$. The average speech spectrum matrix can thus be interpreted as a dictionary of phonemes, which can be composed for producing any arbitrary input speech signal. 

\subsection{Closed-Set Classification}
The last step of the baseline consists of a feature vector computation, and  the actual training of the classifier. For the sake of reproducibility, we use the channel estimate $\hat h$ in  \cref{eq:ideal-speech-estimation} as feature vector for the classification. Moreover, we use a classic \gls{svm} with a \gls{rbf} kernel to perform closed-set classification, and refrain from addressing open-set classification%
\footnote{The extension of the baseline in \cite{mic_class:cuccovillo} to an open-set scenario was addressed in \cite{openset_microphone_classification}, and is compatible with this new proposal to apply denoising as  pre-processing step.}%
.


\section{Denoising Baselines}\label{sec:denoising-baselines}
In the literature, a broad variety of denoising techniques have been proposed to address the denoising problem on signals of different nature for various kinds and levels of noise. In our work, we focus on a selection of well-established DSP-based and more modern AI-based solutions that can be applied to either the time-variant audio signal, or to its time-frequency representation considered as a 2D image.

\subsection{DSP-based: Total Variation}
\begin{subequations}

Image denoising based on total variation has been proposed in~\cite{denoiser:total_variation}. This method is based on the  minimization of a constrained optimization problem.
Specifically, the method aims at finding the image with minimum total variation that minimizes the mean square error with respect to the noisy observation. 

Let us define a noisy image as
\begin{equation}
    v = u + n,
\end{equation}
where $u$ is the ideal noise-free image, and $n$ is an additive noise term.
By denoting the pixel in position $ij$ of image $u$ as $u_{ij}$, we can define the total variation $V(u)$ of the clean image $u$ as the $L_1$ norm of its gradient
\begin{equation}
    V(u)=\sum_{i, j} \sqrt{\abs{u_{i+1, j} -u_{ij}}^2 + \abs{u_{i,j+1} - u_{ij}}^2},
\end{equation}
where $\abs{\cdot}$ denotes the absolute value.

Given a noisy image $v$, image denoising based on total variation consists in estimating the clean image $\hat{u}$ by solving a minimization problem defined as
\begin{equation}
    \hat{u} = \arg\min_u \sum_{i, j} \sqrt{\abs{u_{ij} - v_{ij}}^2} + \lambda V(u),
\end{equation}
where $\lambda$ is a regularization parameter to weight the fidelity term (first term of the equation representing the mean squared error between $u$ and $v$) and the total variation of the estimated $u$ (second term of the equation).
\end{subequations}

\subsection{DSP-based: Non-Local Means}
\begin{subequations}
Denoising based on non-local means is an image denoising algorithm proposed in~\cite{denoiser:nonlocal_means}, which performs the denoising by replacing each pixel by a weighted average of all \emph{similar} pixels in the rest of the image.

Given a noisy picture $v$, the non-local means algorithm retrieves the corresponding clean picture $\hat{u}$ by applying

\begin{equation}
    \hat{u}(i)= \frac{1}{C(i)}
    \sum_{j} w(i,j)v(j), \label{eq:denoising:nonlocal-means}
\end{equation}
where $\hat u(i)$ is the $i$-th pixel of image $\hat u$, $v(j)$ represents the  $j$-th pixel of image $v$, $C(i)$ acts as a normalisation term defined as
\begin{equation}
    C(i) = \sum_{j} w(i,j),
\end{equation}
and the weighting factor $w(i,j)$ depends on the similarity between pixels of $v$ in a neighborhood of its $i$-th pixel and pixels of $v$ in a neighborhood of its $j$-th pixel.
\end{subequations}

\subsection{DSP-based: Bilateral Filtering}
\begin{subequations}
Bilateral filtering is an image denoising algorithm proposed in~\cite{denoiser:bilateral_filtering}, in which each pixel is replaced by a weighted average of \emph{similar, nearby} pixels.

Given a neighborhood $\mathcal{N}_j$ of the $j$-th pixel coordinates of a noisy image $v$, bilateral filtering can be used to recover an estimate of the corresponding clean picture $u$ by applying: 
\begin{equation}
    \hat{u}(i) = \frac{1}{W_p} \int_{j\in\mathcal{N}_j} v(j) \cdot s\left( v(j), v(i) \right) \cdot c\left( j,i \right) \,dj,
\end{equation}
where $s(\cdot)$ is a \emph{similarity} function determining how much the pixel values are alike, $c(\cdot)$ a \emph{closeness} function determining how much the pixel coordinates are close to each other, and $W_p$ a normalization factor.

The authors suggested to use Gaussian functions of the Euclidean distances to determine both similarity and closeness: 
\begin{equation}
    s\left( v(j), v(i) \right) = \exp \frac{1}{2}\left(\frac{\lnorm{ v(j) -v(i) }}{\sigma_s}\right)^2,
\end{equation}
and
\begin{equation}
    c\left( j, i \right) = \exp \frac{1}{2}\left(\frac{\lnorm{ j -i }}{\sigma_c}\right)^2,
\end{equation}
with $\sigma_s$ determining the similarity (photometric) spread, and $\sigma_c$ being the closeness (geometric) spread.
\end{subequations}

\subsection{DSP-based: Wavelet BayesShrink}
\begin{subequations}

The wavelet BayesShrink is an image denoising algorithm proposed in~\cite{denoiser:wavelet_bayesian_shrink}, which leverages wavelet decomposition to filter-out high frequency components associated to the noise.

The authors proposed to focus on the \emph{detail} sub-bands $HH_k$, where $k$ is the scale, computed by applying the two-dimensional dyadic  orthogonal wavelet transform operator $\mathcal{W}$ to the noisy image $v$, yielding the coefficient coefficient matrix
\begin{equation}
    V=\mathcal{W}v.
\end{equation}

In particular, the authors proposed to perform a soft-threshold function to all coefficients $V_{ij}$ of the detail sub-bands, applying 
\begin{equation}
    \hat{U}_{ij}=
    \begin{cases}
    \sgn(V_{ij})\cdot \max\left(\abs{ V_{ij} } -T, 0 \right)), & \text{if } V_{ij}\in HH_k \\
    V_{ij}, & \text{else}
  \end{cases},\label{eq:denoising:wavelet}
\end{equation}
and then transform the image back, thus leaving all low-resolution coefficients unaltered:
\begin{equation}
    \hat{u}=\mathcal{W}^{-1}\hat{U}.
\end{equation}
The core contribution of the paper consisted in defining an \emph{optimal} data-driven threshold $T^\star$ for the shrinking equation in~\eqref{eq:denoising:wavelet}, which minimized the Bayesian risk function associated to the denoising: 
\begin{equation}
    T^\star = \arg \min_T \,E(\hat{U}-U)^2.
\end{equation}
The optimal threshold must be computed per each detailed-subband, and is equal to 
\begin{equation}
    T^\star=\frac{\hat{\sigma}^2}{\hat{\sigma}_U},
\end{equation}
where
\begin{equation}
    \hat{\sigma} = \frac{\text{Median}\left(\abs{ Y_{ij} }\right)}{0.6745}, \,\forall Y_{ij}\in HH_1
\end{equation}
and
\begin{equation}
    \hat{\sigma}_U = \sqrt{ \max\left( \frac{1}{n^2}\sum_{i,j=1}^n  V_{ij}^2 ,0 \right)},
\end{equation}
with $n\times n$ being the size of the $k$-th detailed subband.
\end{subequations}

\subsection{AI-based: Denoising CNN Architecture }
\begin{subequations}

The Denoising CNN architecture (DnCNN) in~\cite{denoiser:dncnn} is a AI-based approach for image denoising aiming at solving the problem by correctly predicting not the original clean image $u$, but rather the \emph{residual} noise $n$. 

The authors start by defining the network as a function $\mathcal{F}(\cdot)$ with parameters $\Theta$, which takes a noisy image $v=u+n$ as input, and performs the following mapping:
\begin{equation}
    \mathcal{F}(v,\Theta) \approx n,
\end{equation}
with $n$ being equal to the (Gaussian white) noise to be  removed.

The loss function $l\left( \cdot \mid \Theta\right)$ to be minimized can thus be defined as the mean squared error between the target noise and the one estimated from the noisy input:
\begin{equation}
    l\left( v, u \mid \Theta\right) = \frac{1}{N}\sum_{i=1}^N\lnorm{ \mathcal{F}\left(v(i), \Theta\right) - (v(i)-u(i)) }^2,
\end{equation}
with $N$ being the amount of pixels in the signal. The advantage of such a procedure, is that a large amount of noise examples can be generated online during training, and thus the network can learn how to predict noises at several \gls{snr} levels -- and potentially more diverse than uniform or Gaussian white noise.

%
%

\end{subequations}

\subsection{AI-based: Audio Denosing Autoencoder}

The Audio Denoising Autoencoder (Audio DAE) in~\cite{denoiser:facebook} is a AI-based approach for audio denoising in the time domain, aiming at reconstructing the clean audio signal $u(t)$ from its noise corrupted version $v(t) = u(t)+n(t)$.

The architecture acts as a denoising autoencoding function $\mathcal{F}(\cdot)$ with parameters $\Theta$, thus realizing the following mapping:
\begin{equation}
    \mathcal{F}(v(t),\Theta) \approx u(t).
\end{equation}
In order to train the network the authors defined the STFT loss $l_{stft}(\Theta)$ as the composition of a \emph{spectral convergence} loss $l_{sc}(\Theta)$ and a \emph{magnitude} loss $l_{mag}(\Theta)$:

\begin{equation}
    l_{stft}\left(u, \hat{u} \mid\Theta\right) =l_{sc}\left(u, \hat{u} \mid\Theta\right) + l_{mag}\left(u, \hat{u} \mid\Theta\right), 
\end{equation}
where
\begin{align}
    l_{sc}\left(u, \hat{u} \mid\Theta\right)   & = \frac{\lnorm{ \abs{STFT(u)} - \abs{STFT(\hat u)}}_F}{\lnorm{\abs{ST FT(u)}}_F}, \\
    l_{mag}\left(u, \hat{u} \mid\Theta\right)  & = \frac{1}{T} \lnorm{ \log \abs{STFT(u)} - \log \abs{STFT(\hat u)} }_1,
\end{align}
with $T$ denoting the input length, $\lnorm{\cdot}_1$ the $L_1$ norm and $\lnorm{\cdot}_F$ the Frobenius norm. 

The peculiarity of the \gls{stft} loss is that it can be computed with several configurations, using different number of \gls{stft} bins, hop sizes and window lengths. Given $M$ such configurations, the authors finally determined the required training loss for the whole architecture: 
\begin{equation}
    l\left( u, \hat{u} \mid \Theta \right) = \frac{1}{T} \left( \lnorm{u-\hat u}_1 + \sum_{m=1}^M l_{stft}^{(m)}\left( u, \hat{u} \mid \Theta \right)\right).
\end{equation}
This composite loss is meant to ensure phase and magnitude coherency of the output audio signal, while at the same time avoiding artifacts which would have appeared by using only one \gls{stft} resolution. 

%
%

\section{Denoiser Selection}\label{sec:denoiser-selection}

In order to determine if and to which extent the denoising baselines outlined in \Cref{sec:denoising-baselines} are applicable for microphone classification, we compared their performances on audio log-powers, which were extracted from the MOBIPHONE~\cite{dataset:mobiphone} dataset according to the overview in \Cref{fig:overview}. 

The MOBIPHONE dataset was collected by recording 10 utterances from 24 speakers, using 21 mobile phones of various models from 7 different brands. After its publication it became a common dataset for benchmarking microphone classification algorithms, and we thus decided to use it both for comparing the denoiser performances, and for running the complete system evaluation in \Cref{sec:evaluation}.

~\\The dataset preparation for the denoiser comparisons was performed as follows:
\begin{enumerate}
    \item We corrupted all audio files in the MOBIPHONE dataset with additive white Gaussian noise, using a \gls{snr} of $25$ dB. 
    \item We extracted log-power spectrograms from the clean MOBIPHONE dataset, obtaining a reference set $X_{ideal}(f)$
    \item We extracted denoised log-power spectrograms from the noisy audio files,  obtaining a benchmark set $\widehat{X}_{25}(f)$
    \item We split both reference set and benchmark set into training and testing portions, obtaining the four distinct sets $X^{train}_{ideal}(f)$, $\widehat{X}^{train}_{25}(f)$,$X^{test}_{ideal}(f)$, $\widehat{X}^{test}_{25}(f)$
\end{enumerate}
{\color{white} A}\\After obtaining the four aforementioned sets, we compared the outcome of the denoising using 3 different metrics: 
\begin{enumerate}
    \item PSNR: Average \gls{psnr} between corresponding pairs of $X^{test}_{ideal}(f)$ and $\widehat{X}^{test}_{25}(f)$
    \item SSIM: Average \gls{ssim} between corresponding pairs of $X^{test}_{ideal}(f)$ and $\widehat{X}^{test}_{25}(f)$
    \item MCA: \gls{mca} of the baseline trained on $X^{train}_{ideal}(f)$ and tested on $\widehat{X}^{test}_{25}(f)$
\end{enumerate}
The first two metrics relate directly to the visual quality of the denoising: The \gls{psnr} quantifies the closeness in terms of pixel energy between the original spectrogram and the denoised one, while the \gls{ssim} their similarity in terms of luminance, contrast and structure. The \gls{mca} metric is meant to capture to which extent the classification is possible after the denoising operation: Aggressive denoisers may remove too much content from the log-power spectrogram, while ineffective ones may remove too little disturbance for being of any help. 

\begin{figure*}[ht]
  \begin{minipage}[b]{\specgramminipagewidth}
    \centering
    \includegraphics[width=\specgramwidth]{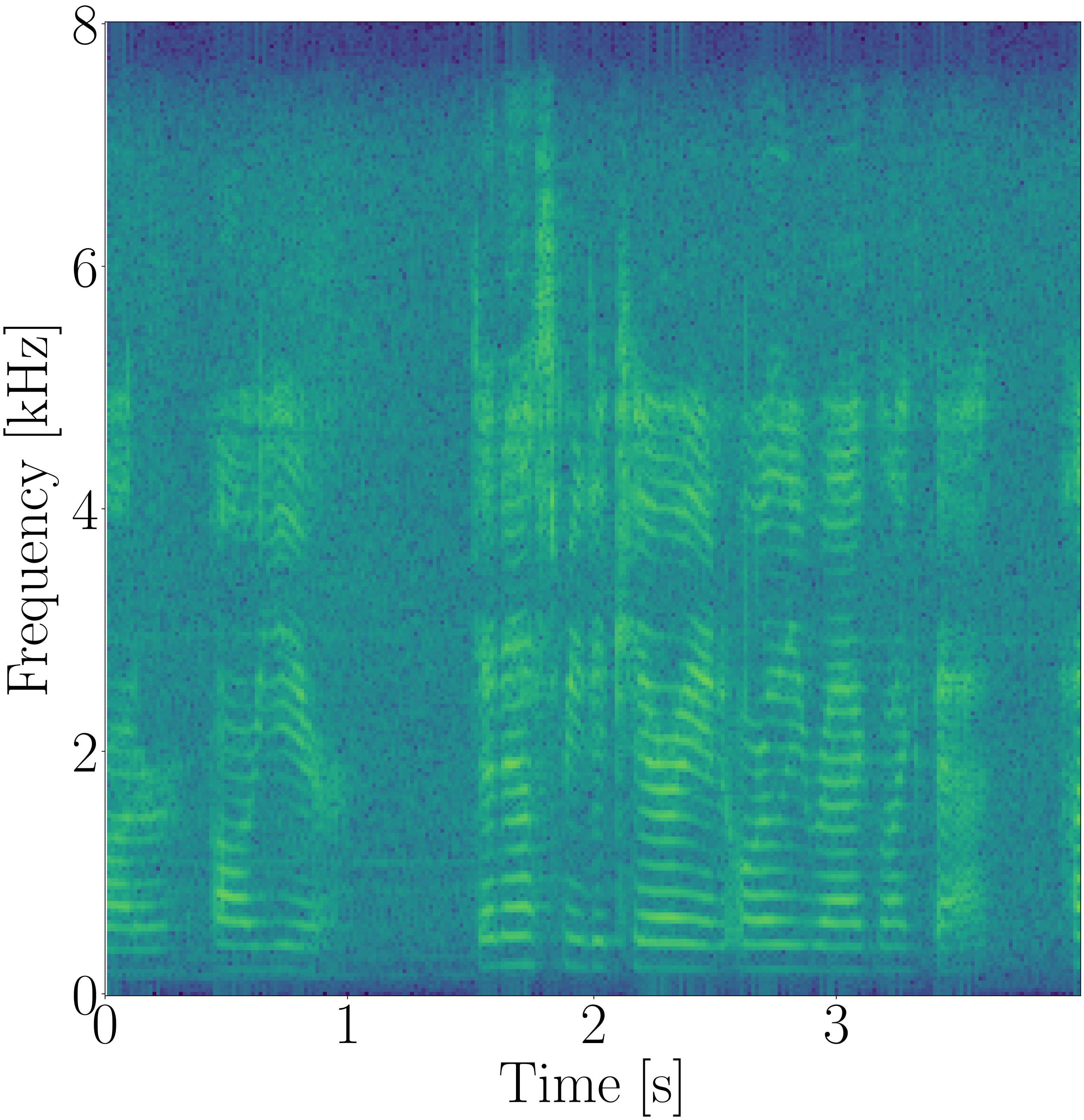}
    \subcaption{Original Specgram}
  \end{minipage}%
  \begin{minipage}[b]{\specgramminipagewidth}
    \centering
    \includegraphics[width=\specgramwidth]{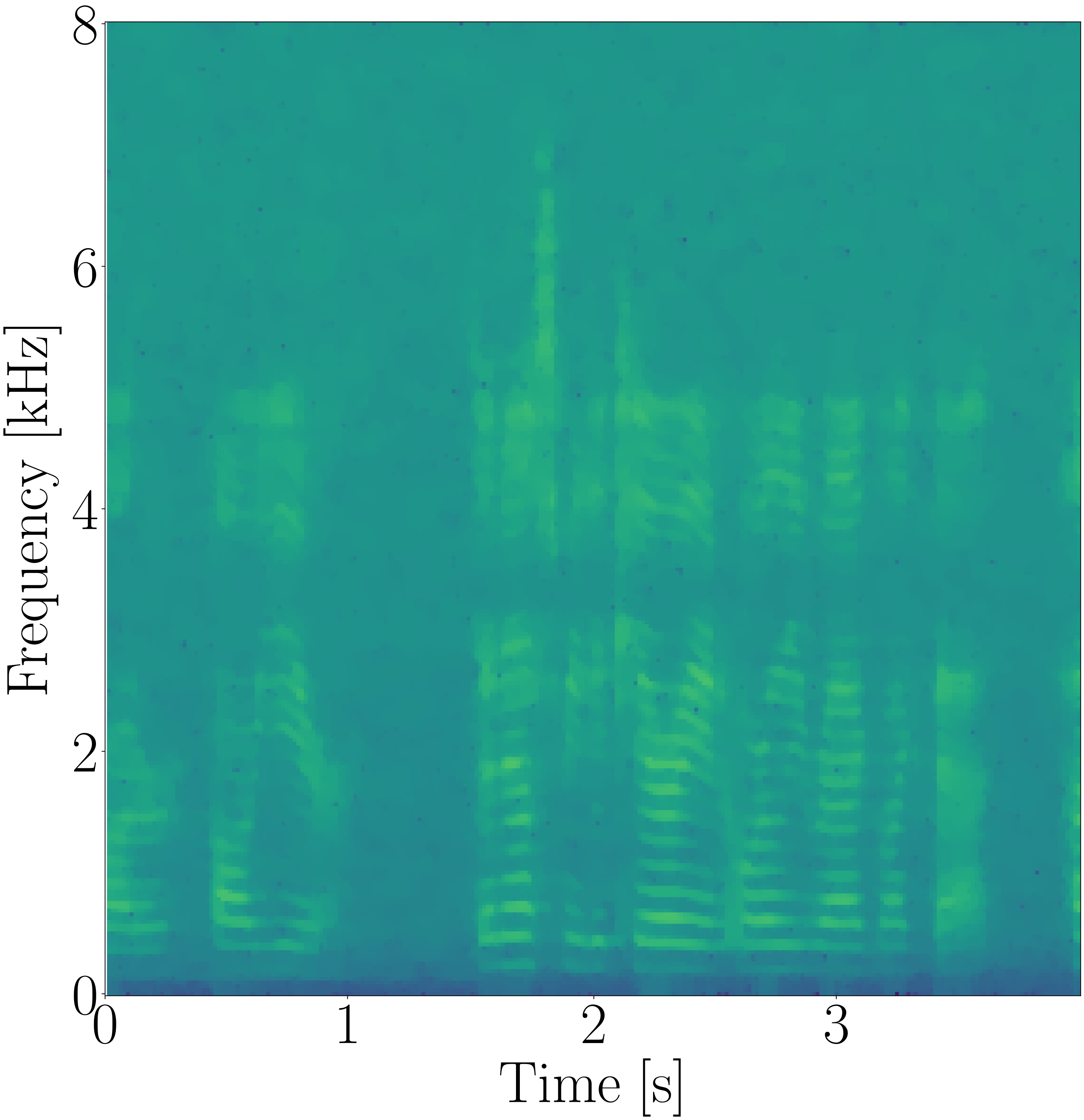}
    \subcaption{Total Variation~\cite{denoiser:total_variation}}
  \end{minipage}%
  \begin{minipage}[b]{\specgramminipagewidth}
    \centering
    \includegraphics[width=\specgramwidth]{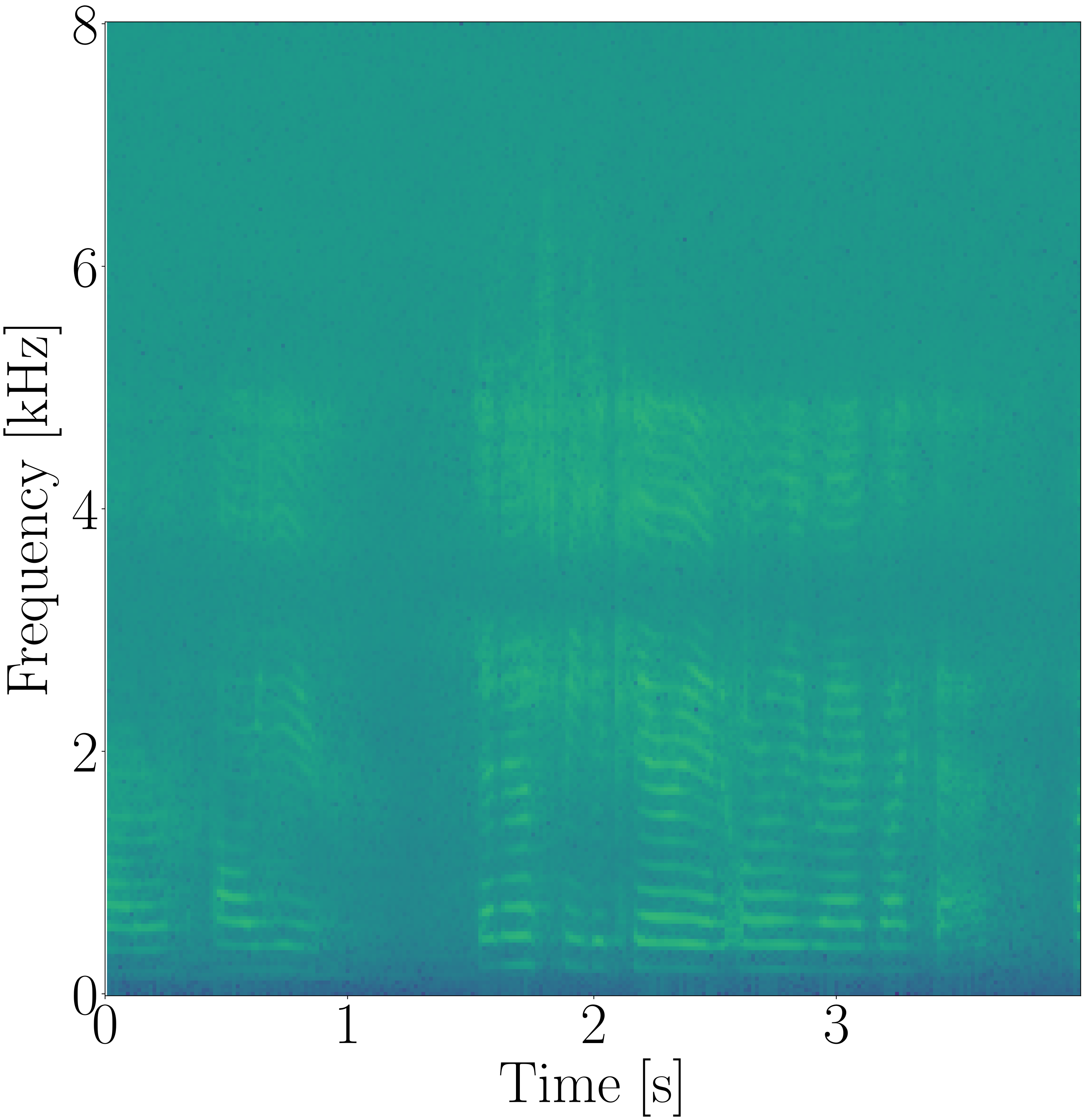}
    \subcaption{Bilateral Filtering~\cite{denoiser:bilateral_filtering}}
  \end{minipage}%
  \begin{minipage}[b]{\specgramminipagewidth}
    \centering
    \includegraphics[width=\specgramwidth]{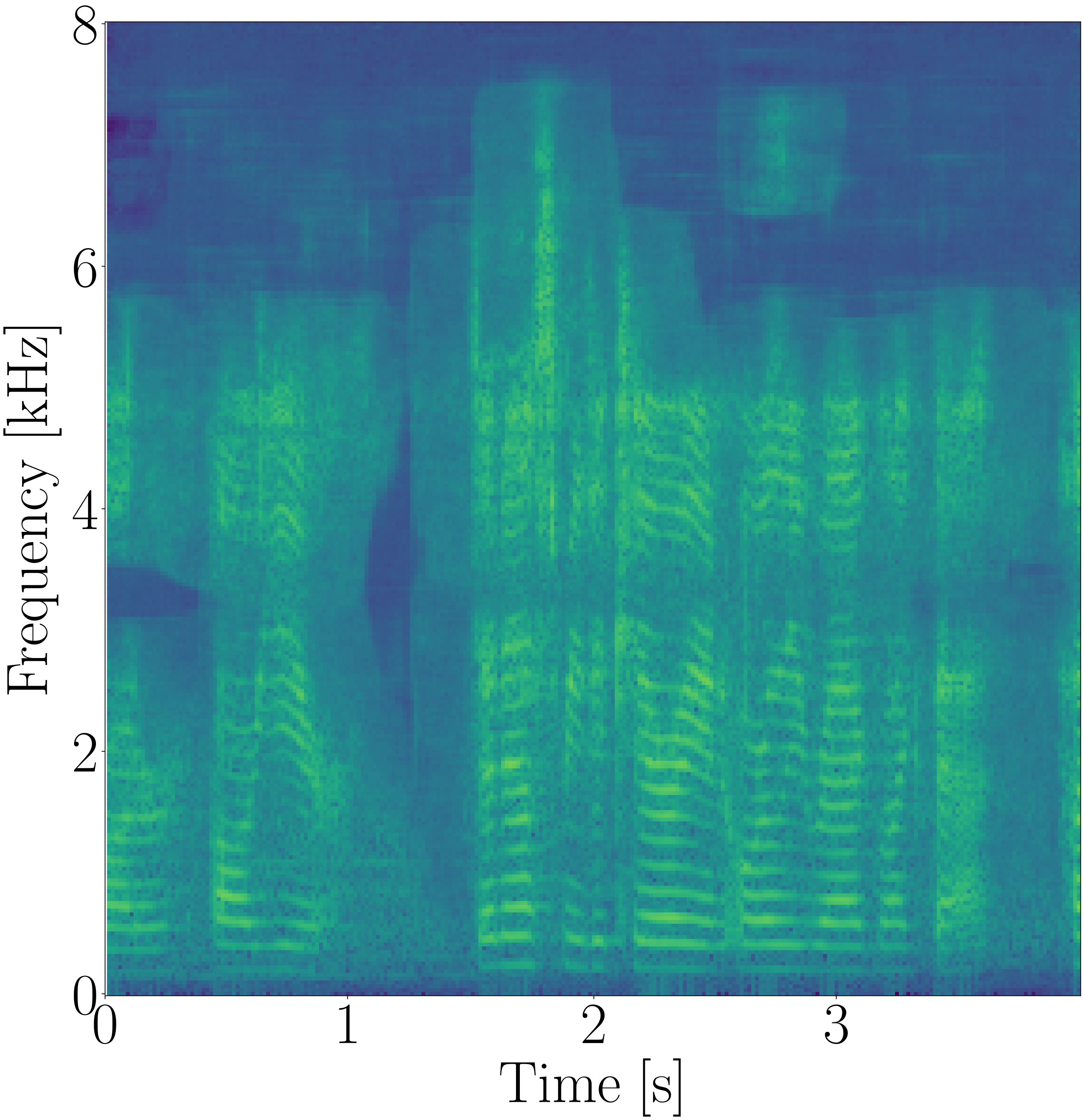}
    \subcaption{DnCNN Architecture~\cite{denoiser:dncnn}}
  \end{minipage}\\[1.2em]
  \begin{minipage}[b]{\specgramminipagewidth}
    \centering
    \includegraphics[width=\specgramwidth]{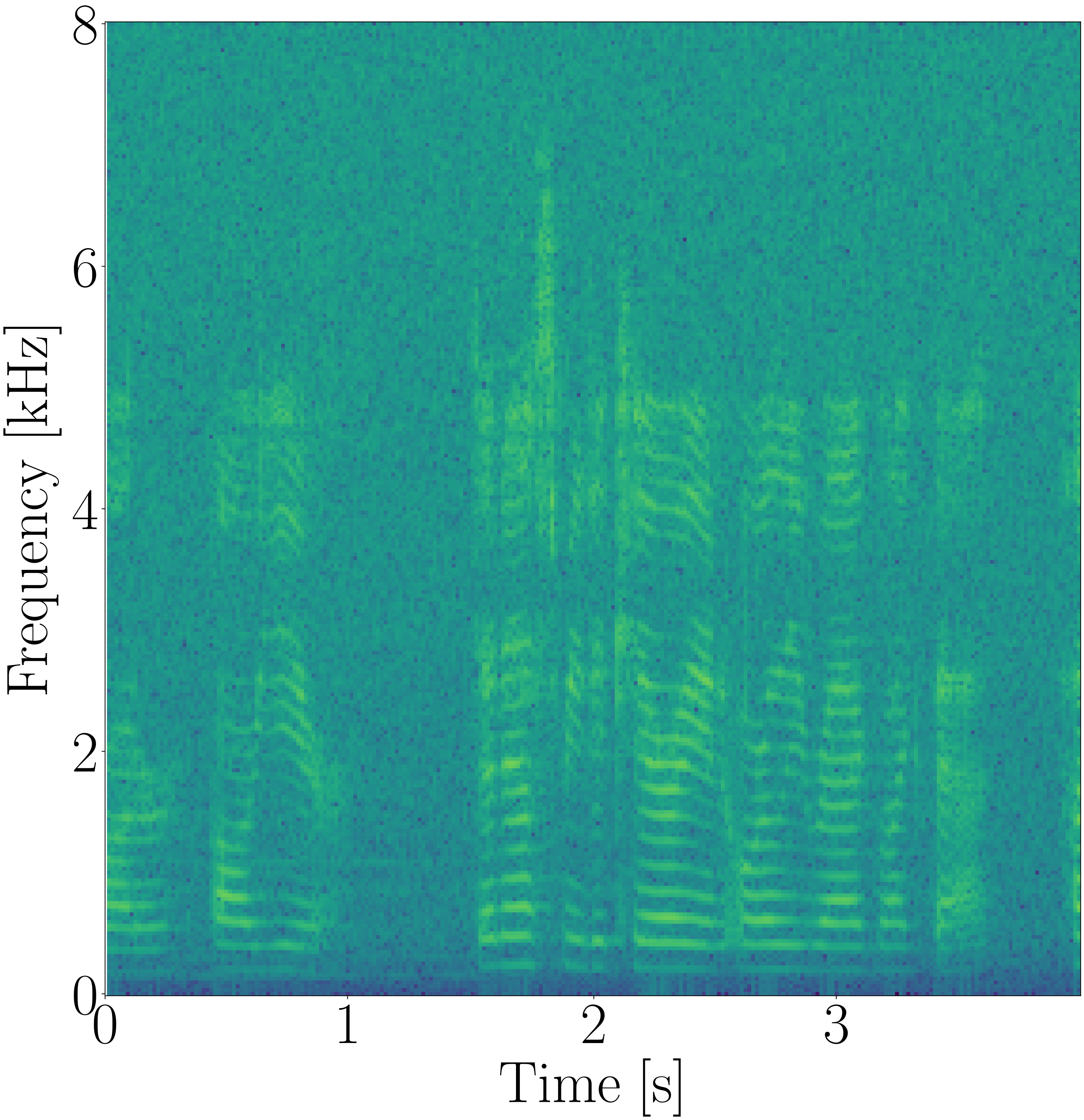}
    \subcaption{Noisy Specgram}
  \end{minipage}%
  \begin{minipage}[b]{\specgramminipagewidth}
    \centering
    \includegraphics[width=\specgramwidth]{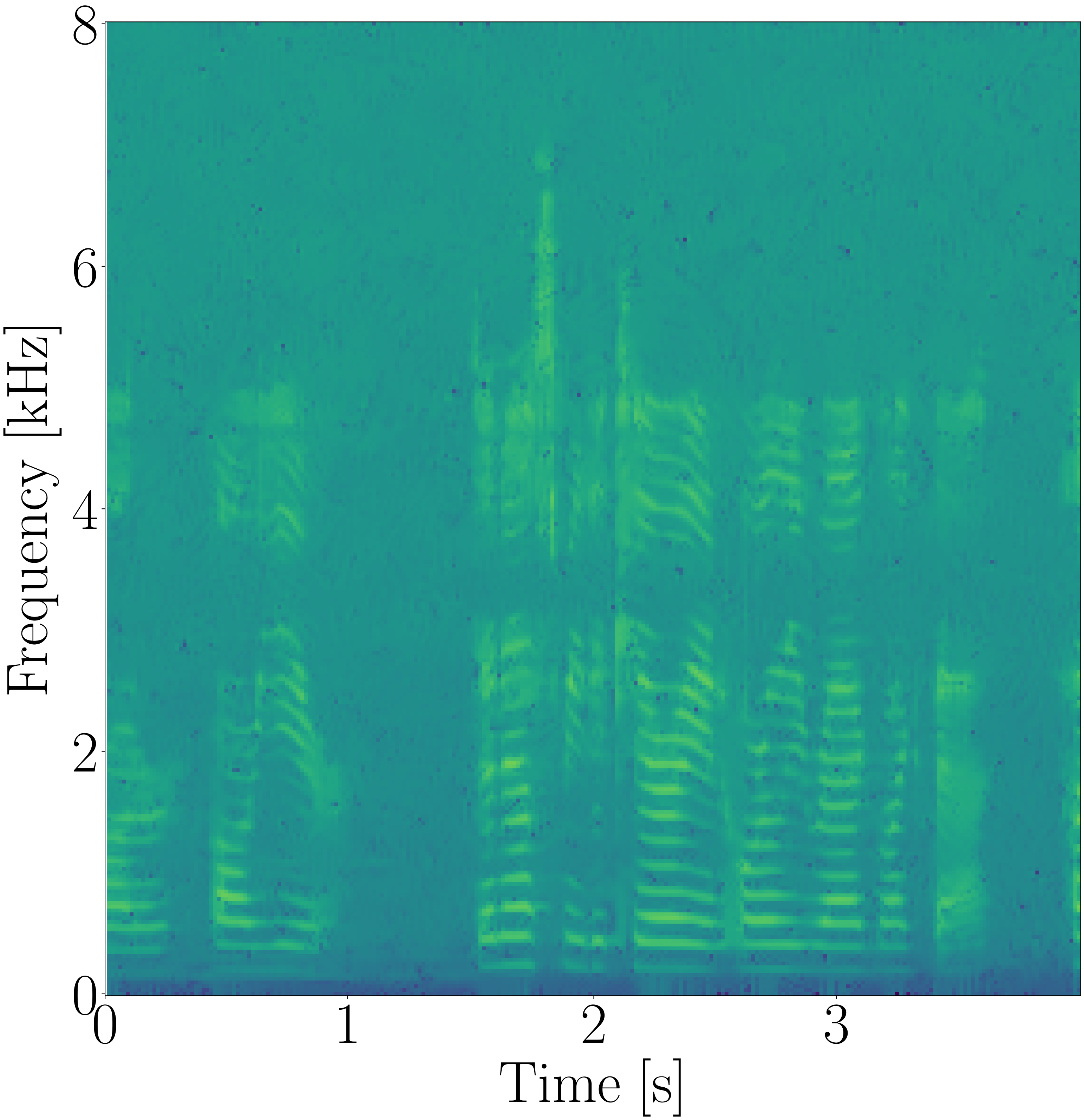}
    \subcaption{Non-Local Means~\cite{denoiser:nonlocal_means}}
  \end{minipage}%
  \begin{minipage}[b]{\specgramminipagewidth}
    \centering
    \includegraphics[width=\specgramwidth]{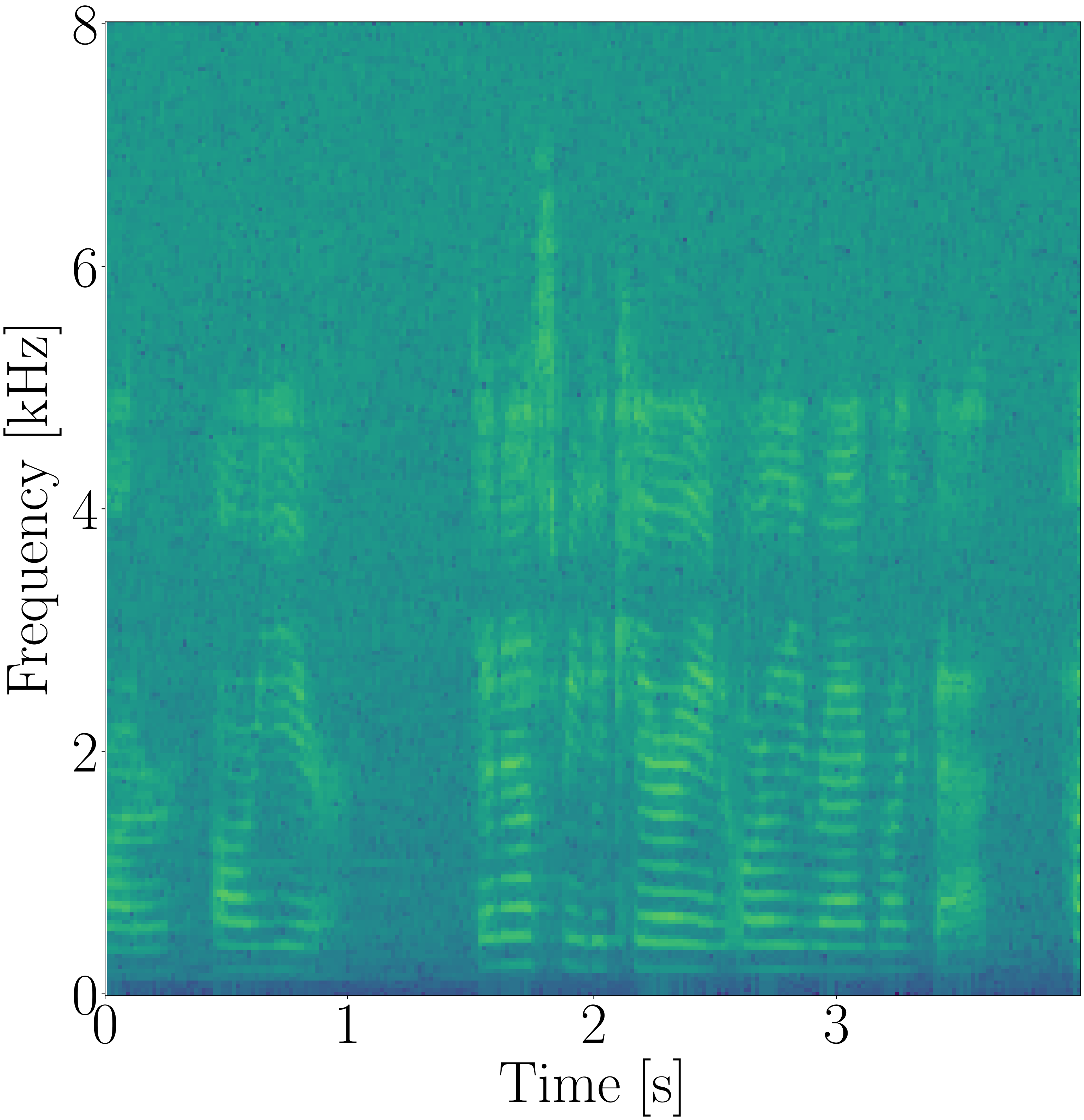}
    \subcaption{Wavelet BayesShrink~\cite{denoiser:wavelet_bayesian_shrink}}
  \end{minipage}%
  \begin{minipage}[b]{\specgramminipagewidth}
    \centering
    \includegraphics[width=\specgramwidth]{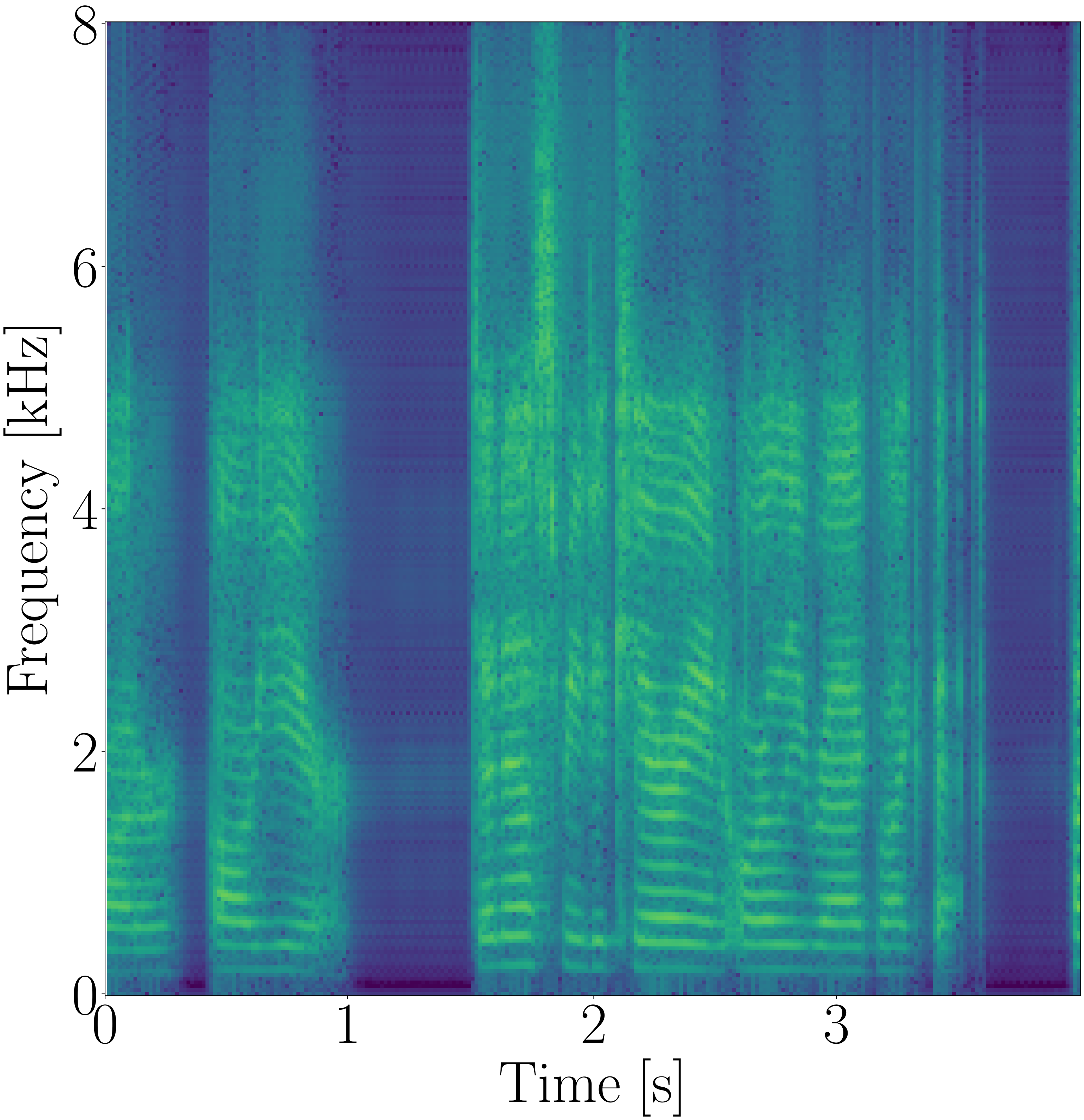}
    \subcaption{Audio DAE Architecture~\cite{denoiser:facebook}}
  \end{minipage}
  \caption{Example denoising of log-power spectrograms, with audio SNR=25dB }\label{fig:denoised-specgrams}
\end{figure*}

~\\
The outcome of this evaluation is reported in \Cref{tab:denoiser-selection}, with example denoised spectrograms being depicted in \Cref{fig:denoised-specgrams}.

\begin{table}[h!b] 
\centering
\caption{Denoising Baseline Benchmarks}
\label{tab:denoiser-selection}
\begin{tabular}{@{}lllll@{}}
\toprule
\multirow{2}{*}{Denoiser} && \multicolumn{3}{c}{Performances (\#)}       \\ \cmidrule(l){3-5} 
                                                            && PSNR & SSIM & MCA \\ \midrule
Total Variation~\cite{denoiser:total_variation}             && 20.56   & 0.81     & 34.40 \\
Non-Local Means~\cite{denoiser:nonlocal_means}              && 20.59   & 0.83     & 36.69  \\
Bilater Filtering~\cite{denoiser:bilateral_filtering}       && 20.33   & 0.81     & 33.67 \\
Wavelet BayesShrink~\cite{denoiser:wavelet_bayesian_shrink} && 20.62   & 0.82     & 39.06 \\
DnCNN Architecture~\cite{denoiser:dncnn}                    && \textbf{27.80}   & \textbf{0.86}    & \textbf{69.09} \\
Audio DAE Architecture~\cite{denoiser:facebook}             && 21.02   & 0.76     & 26.47 \\\bottomrule
\end{tabular}
\end{table}

We can observe that the four \gls{dsp}-based denoising algorithms perform relatively on-par in terms of both visual quality and microphone accuracy. The latter, however, is not sufficient for practical applications, given the \gls{mca} score below 40\%. 

The Audio DAE is superior to \gls{dsp}-based denoisers in terms of \gls{psnr}, but achieves a lower \gls{ssim}. The reason for this counter-intuitive behavior can be identified by checking \Cref{fig:denoised-specgrams}: The Audio DAE is pretty aggressive within the silent portions of the audio signal, and generates some edges in the spectrogram that lower the \gls{ssim}. The same aggressiveness is probably also the cause for a further decrease of the \gls{mca} score, which is reduced to about 26\%. 

The DnCNN Architecture seems to be the most promising alternative: Its \gls{psnr} is superior to the Audio DAE by more than 6dB, and the \gls{ssim} is beyond the one achieved by the \gls{dsp}-base denoisers. If we consider the examples in \Cref{fig:denoised-specgrams}, the higher \gls{ssim} is probably due to the speech components being sharper than in other spectrograms. The most promising score, however, is the \gls{mca} itself: Without retraining the classifier, which would probably improve the performances but could be a costly operation, we achieved an accuracy of about 69\%, which is significantly beyond the accuracy of all other alternatives. We hence selected the DnCNN Architecture to perform the complete system evaluation, which is outlined in the next section.

\section{Evaluation}\label{sec:evaluation}
The final evaluation of our proposed integrated approach, combining AI-based denoising performed by a DnCNN architecture~\cite{denoiser:dncnn} and microphone classification based on blind channel estimation~\cite{mic_class:cuccovillo,openset_microphone_classification}, involved several datasets in conjunction.

Similarly as for the denoiser selection, the core evaluation dataset was the MOBIPHONE dataset, which was resampled to 16 kHz and split in segments of 4.096 seconds. This segment length, in conjunction with a \gls{stft} using 512 points and 50\% overlap, led us to a set of 180 log-power spectrogram examples of size $256\times256$ per each of the $21$ classes in the dataset. Furthermore, we split training and testing examples according to a speaker-wise logic: Spectrograms related to the utterances of the first 19 speakers were used for training (142 examples per class), and the remaining ones for testing (38 examples per class).

In addition to the original MOBIPHONE dataset, we created noise-corrupted versions of both the train and the test examples, using audio \glspl{snr} equals to 20,35,30 and 35 dB. If we assume that the initial recordings had infinite SNR, each set (and its denoised equivalent) can be uniquely identified as follows:
\begin{align*}
    X^{train}_{snr}&: \text{Noisy (SNR=$snr$) training examples}\\
    X^{test}_{snr}&: \text{Noisy (SNR=$snr$) test examples}\\
    \widehat{X}^{train}_{snr}&:\text{Denoised (from SNR=$snr$) training examples}\\
    \widehat{X}^{test}_{snr}&: \text{Denoised (from SNR=$snr$) test examples}
\end{align*}

The \gls{svm} used for microphone classification was trained \emph{once} on the $X^{train}_{\infty}$ set, selecting a \gls{rbf} kernel with $\gamma=1/(256\,\sigma^2_x)$ --where $\sigma^2_x$ is equal to the variance of the normalized feature vectors--, and then tested separately on each $\widehat{X}^{test}_{snr}$. The absence of retraining has the advantage of simulating the behavior of \emph{pre-existing} classification pipelines which are exposed to noisy content. 
The DnCNN used for spectrogram denoising was instead trained on each available  $\left( X^{train}_{snr}, X^{train}_{\infty}\right)$ pair. Due to to the speaker-wise split, test examples where thus unseen for both the denoiser and the classifier.   

The \gls{gmm} included in the microphone classification baseline was trained by combining utterances from the LibriSpeech corpus~\cite{dataset:librispeech}, until reaching a total duration of one hour. According to  \cite{mic_class:cuccovillo}, we trained a \gls{gmm} with 1024 mixtures, using 12 \glspl{mfcc} per frame, computed for each frame of the aforementioned denoised \gls{stft}. Being the speakers in the LibriSpeech corpus absent from the MOBIPHONE dataset -- which includes utterances from TIMIT~\cite{dataset:timit} -- we tried to ensure, once again, that examples were completely unknown to the integrated classification algorithm. 

The final outcome of the evaluation is presented in \Cref{tab:results:without-denoising} and \Cref{tab:results:with-denoising}. 
The results in \Cref{tab:results:without-denoising} describe the performance of the system if applied without any precautions to noisy content: From being nearly perfect -- with accuracy, precision and recall all higher than 99\% -- all metrics drop by more than half if the quality decrease to a \gls{snr} of 30 dB of lower. Even a very light noise with 35 dB \gls{snr} is enough to lower the accuracy to about 60\%. The loss of performances was not unexpected, due to the lack of any noise term in the channel estimation formulation of \cref{eq:channel-estimation}, but is still higher than we would have imagined, and confirms the need for countermeasures whenever analysing noisy input audio files. 

\begin{table}[hb]
\centering
\caption{Results \emph{without} including denoising}\label{tab:results:without-denoising}
\begin{tabular}{@{}cllll@{}}
\toprule
\multirow{2}{*}{SNR (dB)} &  & \multicolumn{3}{c}{Performances (\%)}       \\ \cmidrule(l){3-5} 
                          &  & Accuracy & Precision & Recall \\ \midrule
+$\infty$                 &  & 99.21   & 0.992  & 0.992 \\
35                        &  & 60.65   & 0.643  & 0.603                        \\
30                        &  & 50.36   & 0.514  & 0.500                        \\
25                        &  & 41.81   & 0.437  & 0.415                        \\
20                        &  & 36.05   & 0.409  & 0.353                         \\ \bottomrule
\end{tabular}
\end{table}

The results in \Cref{tab:results:with-denoising}, instead, describe the performance of the system  we proposed, in which the denoising is active and the classifier is not retrained nor aware of the denoising taking place. We see a very consistent behavior in performances for all noise-corrupted cases: Accuracy, precision and recall, which dropped severely without countermeasures, raise by about 25\% compared to the baseline, and at least for low-noise scenarios -- \gls{snr} of at least 30 dB -- they consistently score beyond 80\%. There is a perceivable decrease compared to the initial 99\% of accuracy, but the system can still provide useful results. Analysis performances for clean content decrease slightly to about 95\%, i.e., the denoising procedure is not completely transparent, but not to a prohibitive point. We will discuss a mitigation strategy for this issue and other further research directions in the following section. 

\begin{table}[hb]
\centering
\caption{Results of our proposed approach}\label{tab:results:with-denoising}
\begin{tabular}{@{}cllll@{}}
\toprule
\multirow{2}{*}{SNR (dB)} &  & \multicolumn{3}{c}{Performances (\%)}       \\ \cmidrule(l){3-5} 
                          &  & Accuracy & Precision & Recall \\ \midrule
+$\infty$                 &  & 95.11   & 0.955   & 0.950 \\
35                        &  & 84.17   & 0.872   & 0.841 \\
30                        &  & 83.71   & 0.861   & 0.839 \\          
25                        &  & 69.09   & 0.771   & 0.694 \\ 
20                        &  & 57.54   & 0.695   & 0.578 \\          \bottomrule\\
\end{tabular}
\end{table}

\section{Conclusions and Outlook}\label{sec:outlook}
To our knowledge, this work represents the first systematic approach addressing the influence of noise on microphone classification. The proposed pipeline was tested on a widely accepted benchmark dataset, and proved to be effective in lessening the negative influence of noise on the classification task: The proposed algorithm achieved an accuracy beyond 80\% for noisy conditions with audio \gls{snr} of at least 30dB, providing an accuracy increase of about 25\% in comparison with the initial pipeline, \emph{without} the need for re-training the classifier under analysis.

In a parallel submission~\cite{giganti_2022_arXiv}, we will address two additional research questions which we could not cover within this pages: Testing whether DnCNN-based denoising can be applied successfully to other state-of-the-art feature vectors for microphone classification, and investigating whether data augmentation can boost the overall accuracy and avoid the slight performance drops for clean-yet-denoised test files.

Furthermore, we plan to verify the efficacy of the DnCNN denoising with alternative classification algorithms, and investigate countermeasures to more complex noises than the uniform white Gaussian noise addressed by this publication, e.g., babble noise, car noise or background music.

We will also investigate whether a similar approach could be used to lessen the negative influence of lossy compression and transcoding, which are also likely to occur on social networks. Finally, we plan to experiment with new denoising techniques which are likely to emerge within the image-processing domain.

\begin{acks}
This paper was supported by the 
\grantsponsor{AI4Media}{EU H2020 AI4Media research project}{https://www.ai4media.eu/} (grant no. \grantnum{AI4Media}{951911}). 
\end{acks}
\bibliographystyle{ACM-Reference-Format}
\bibliography{references}



\end{document}